\documentclass[%
 reprint,
 amsmath,amssymb,
 aps,
]{revtex4-2}

\usepackage[utf8]{inputenc}
\usepackage{physics}
\usepackage{graphicx}
\usepackage{amsmath}
\usepackage{amsthm}
\usepackage{amsfonts}
\usepackage{float}
\usepackage{braket}
\usepackage{url}
\usepackage[colorlinks=true, pdfstartview=FitV, linkcolor=red, citecolor=blue, urlcolor=blue]{hyperref}
\usepackage{slashed}
\usepackage[normalem]{ulem}
\usepackage{float}
\usepackage{braket}

\graphicspath{{./figures/}}

\theoremstyle{definition}

\newtheorem*{thm*}{Theorem}

\newtheorem*{defn*}{Definition}

\newtheorem*{lem*}{Lemma}

\newtheorem*{rem*}{Remark}

\newtheorem*{con*}{Conjecture}

\newtheorem*{cor*}{Corollary}

\newtheorem*{prop*}{Proposition}

\newtheorem*{hypoth*}{Hypothesis}

\newtheorem*{claim*}{Claim}

\begin{document} 
\title{Hyperbolic Band Theory under Magnetic Field and Dirac Cones on a Higher Genus Surface}



\author{Kazuki Ikeda}
\email{kazuki7131@gmail.com}
\affiliation{Department of Mathematics and Statistics
$\&$ Centre for Quantum Topology and Its Applications (quanTA), University of Saskatchewan, Saskatoon, Saskatchewan S7N 5E6, Canada}
\author{Shoto Aoki}
\affiliation{Department of Physics, Osaka University, Toyonaka, Osaka 5600043, Japan}
\author{Yoshiyuki Matsuki}
\affiliation{Department of Physics, Osaka University, Toyonaka, Osaka 5600043, Japan}

\begin{abstract}We explore the hyperbolic band theory under a magnetic field for the first time. Our theory is a general extension of the conventional band theory defined on a Euclidean lattice into the band theory on a general hyperbolic lattice/Riemann surface. Our methods and results can be confirmed experimentally by circuit quantum electrodynamics (cQED), which enables us to create novel materials in a hyperbolic space. To investigate the band structures, we construct directly the hyperbolic magnetic Bloch states and find that they form Dirac cones on a coordinate neighborhood, by which they can be regarded as a global quantum gravity solution detectable in a laboratory. Besides this is the first explicit example of a massless Dirac state on a higher genus surface. Moreover we show that the energy spectrum exhibits an unusual fractal structure refracting the negative curvature, when plotted as a function of a magnetic flux.
\end{abstract}

\maketitle 

\section{Introduction}
Circuit quantum electrodynamics (cQED) is an architecture
for studying quantum information and quantum optics, in which artificial atoms/superconducting qubits are coupled to photon in a one-dimensional on-chip resonator \cite{2004PhRvA..69f2320B,2017PhRvX...7a1016F,https://doi.org/10.1002/andp.201200261,PhysRevB.94.014506,2008JAP...104k3904G,2012NatPh...8..292H}. The field of cQED provides a platform for quantum simulation of quantum many-body systems, and is a promising candidate for universal quantum computation. As those references show, quantum simulation with cQED has been considered for materials in the Euclidean geometry for more than a decade. More recently, a new experimental realization of hyperbolic lattices in cQED was presented~\cite{2019Natur.571...45K}. This seminal work opens up a new direction for quantum simulation of materials in a curved space. Implementation of holographic error correction codes \cite{2019PhRvR...1c3079J,2015JHEP...06..149P} is of a practical interest, not only for high energy physics but also for universal quantum computation. In fact a hyperbolic space, which has some negative curvature, cannot be isometrically embedded in the Euclidean space. Therefore there have been some technical difficulties in experimental study on materials in a hyperbolic space. So developing such a geometrically deformable quantum simulator endows us with a significant advantage in studying quantum physics in a curved space-time. In particular understanding quantum mechanical phenomena in a non-Euclidean geometry in the presence of gravity is one of the most important challenges for the modern physicists. Furthermore information processing on a hyperbolic lattice has some practical and crucial meanings for information technology \cite{2010PhRvE..82c6106K,2010NatCo...1...62B}, where it is argued that the connectivity of the internet is given by a hyperbolic map. 

\begin{figure*}
\begin{minipage}{0.49\hsize}
        \centering
    \includegraphics[width=\hsize, bb=0 0 400 400]{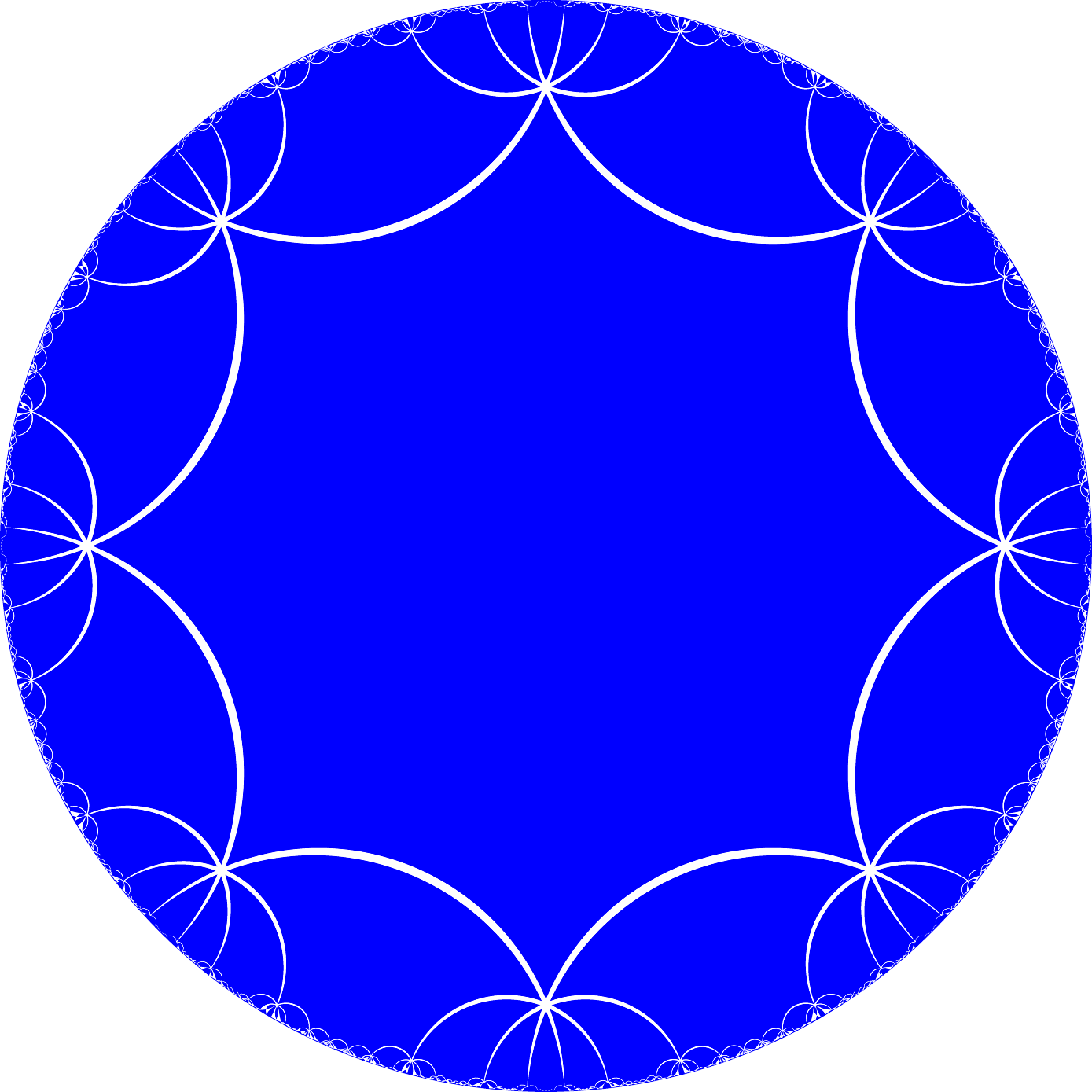}
\end{minipage}
\begin{minipage}{0.49\hsize}
        \centering
    \includegraphics[width=\hsize, bb=0 0 819 515]{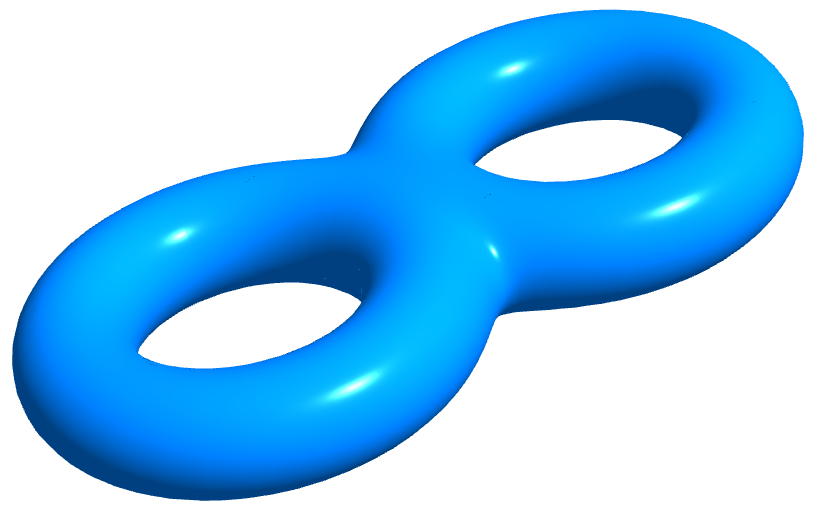}
\end{minipage}
    \caption{The regular tessellation $\{8,8\}$ of the hyperbolic plane, which gives the genus-2 surface.}
    \label{fig:lattice}
\end{figure*}

Motivated by this novel experimental environment of quantum simulation with cQED, we construct the first generalization of Bloch band theory on hyperbolic lattices under a magnetic field perpendicular to the system (Fig.~\ref{fig:lattice}). We aim at creating and investigating a theory of such a new class of materials, based on Riemann surface theory and algebraic geometry. In fact, the pairwise identification of sides of a unit cell of a hyperbolic $4g$-gon with the Fuchsian group gives a compact Riemann surface of genus $g\ge2$. Band theory on a hyperbolic lattices without a magnetic field was studied in \cite{COMTET1987185,doi:10.1063/1.526781} and recently developed in \cite{2020arXiv200805489M} more formally. So our theory is a natural, general and formal extension of the conventional band theory with Bloch states on a Euclidean lattice into that on a hyperbolic lattice. For any Hamiltonian with the $\{4g, 4g\}$ tessellation symmetry of the hyperbolic plane, we derive eigenstates associated with the magnetic Bloch conditions under the magnetic Fuchsian group of the tessellation, which is discrete but non-commutative. The magnetic Bloch states are generalized to automorphic functions. The construction of such a band theory on a hyperbolic lattice with a constant negative curvature is extremely non-trivial due to the absence of commutative translation symmetries. In fact, imposing the magnetic Bloch condition is not straightforwardly applicable to the hyperbolic band theory. To circumvent such a technical difficulty, we create the notion of magnetic Fuchsian group so that the Hamiltonian becomes simultaneously diagonalizable with the magnetic translations. Those mathematical and technical subjects are summarized in \cite{2021arXiv210710586A}. Although the Bloch wavefunctions were first introduced to describe the propagation of electrons in crystalline solids, this phenomenon applies to a generic propagation of wave-like phenomenon in a periodic system, including atomic matter waves in optical lattices, light in photonic crystals, and sound in acoustic systems. In fact the Bloch theorem states that solutions of the Schr\"{o}dinger equation in a periodic potential correspond to plane waves. The periodicity of the underlying system is crucial for the existence of a Bloch state. In our case, the tessellation symmetry and the magnetic Fuchsian group guarantee this condition.  

This study will contribute to a wide range of fields, including material design, condensed matter physics, quantum computation/simulation, high energy physics and mathematical physics. The rest of this article is organized as follows. In Sec.~\ref{sec:setup} we describe our system and prepare the Hamiltonian. In Sec.~\ref{sec:band} we present our numerical results of band structures under a magnetic field. We report that they form Dirac cones, which implies that there are massless states obeying the Dirac equation on a coordinate neighborhood. This is the first explicit example of the Dirac states on a higher genus surface. We may link them to a theory of quantum gravity since they live in a hyperbolic space, which is a time slice of the three dimensional Anti-de Sitter $(\text{AdS}_3)$ space, a solution of the Einstein equation. Since it is a massless state, we expect it to travel the hyperbolic plane as fast as light. Extracting some gravitational information will enhance our understanding on a quantum gravity theory. In Sec.~\ref{sec:butt} we present that the energy spectrum exhibits an unusual fractal pattern, where each stack of bands forms a Hofstadter butterfly. The existence of many stacks of butterflies clearly distinguishes the hyperbolic band theory from the conventional Euclidean band theory. In Sec.~\ref{sec:hall}, we describe the Brilluoin zone and the Hall Conductance. 

\section{\label{sec:setup}Hamiltonian on Hyperbolic Surface under Magnetic Field}
 Here we show our band theory on a hyperbolic surface under a  magnetic field perpendicular to the system. In the succeeding sections, we analyze the band structure (Fig.~\ref{fig:band}) and show that they exhibit the fractal structure (Fig.~\ref{fig:the spectrum of H lattice}). Let $\mathbb{H}=\Set{z=x+iy \in \mathbb{C}| y>0}$ be the upper half-plane equipped with its usual Poincar\'{e} metric $ds^2=\frac{dx^2+dy^2}{y^2}$. Let $A\in\Omega^1(\mathbb{H})$ be a one-form on $\mathbb{H}$. Using the Laplace-Beltrami operator, we can write the Hamiltonian with a magnetic field as 
\begin{align}
\begin{aligned}
\label{eq:Hamiltonian}
H&=\frac{1}{2m}\frac{1}{\sqrt{g}}(p_\mu-A_\mu)\sqrt{g}g^{\mu\nu}(p_{\nu}-A_\nu)\\
&=\frac{1}{2m}y^2((p_x-A_x)^2+(p_y-A_y)^2),
\end{aligned}
\end{align}
where $p_\mu=-i\partial_\mu$. Since we are interested in the quantum Hall effect, we take the constant magnetic field that can be written as $dA=B\omega$, where $\omega=y^{-2}dx\wedge dy$ is the area element and $B\in\mathbb{R}$ is some fixed value. We use $A=\frac{B}{y}dx$, then the Hamiltonian becomes 
\begin{align}
\begin{aligned}
    H&=\frac{y^2}{2m}\left(\left(p_x-\frac{B}{y}\right)^2+p_y^2\right)\\
    &\label{eq:V}=-\Delta+i\frac{B}{m}y\partial_x+\frac{B^2}{2m}. 
\end{aligned}
\end{align}

In what follows, we show a general procedure to address electron states in a hyperbolic surface under a uniform magnetic field. We address electron states on a lattice, so-called a Poincar\'{e} tile. Note that in a Euclidean lattice, the tight-binding Hamiltonian commutes with the translation operators. Therefore the Bloch states are their simultaneous eigenstates. In fact, the translation operators are commutative in the absence of a magnetic field. However in a hyperbolic lattice, the Fuchsian group is non-commutative. Therefore the construction of the magnetic Bloch state is quite non-trivial. Nevertheless, we verify their existence and give their concrete forms. 

Our prescription for obtaining the band structure on a hyperbolic tiling is as follows:
\begin{enumerate}
    \item Find a lattice Hamiltonian $H_{\text{lattice}}$ that converges into the original Hamiltonian as the lattice parameter goes to zero. 
    \item Deform the Hamiltonian $H_\text{lattice}$ in such a way that it respects the original tiling $\{4g,4g\}$.
    \item Diagonalize $H_{\text{lattice}}$
\end{enumerate}
In \cite{2021arXiv210710586A}, we verify that our following arguments are mathematically valid and respect all of the procedure above. In what follows we demonstrate how this protocol works for the $g=2$ case. We define operators 
\begin{align}
\begin{aligned}\label{eq:generator of magnetic SL2}
    \hat{S}_B&=(1+x^2-y^2)\pdv{}{x}+2xy\pdv{}{y}+2iBy \\
    \hat{T}_B&=\pdv{}{x}\\
    \hat{U}_B&=2x\pdv{}{x}+2y\pdv{}{y}
\end{aligned}
\end{align}
and write 
\begin{equation}
    \hat{J}_1=\frac{1}{2}(2\hat{T}_B-\hat{S}_B),~\hat{J}_2=\frac{i}{2}\hat{S}_B,~\hat{J}_3=\frac{1}{2}\hat{U}_B,
\end{equation}
which satisfy the commutation relation 
\begin{equation}
    [\hat{J}_i,\hat{J}_j]=i\epsilon_{ijk} \hat{J}_k. 
\end{equation}
Using them, we can rewrite the Hamiltonian as 
\begin{equation}
\label{eq:Ham}
    H=\frac{1}{2m}\left(-\sum_{i=1}^3 \hat{J}^2_i+B^2\right). 
\end{equation}
Then we define the lattice Hamiltonian $H_\text{lattice}$ as 
\begin{equation}
\label{eq:Ham_latt}
    H_\text{lattice}=-\sum_{i=1}^3\frac{1}{a^2_i}\left(\exp(a_i \hat{J}_i)+\exp(-a_i\hat{J}_i)\right), 
\end{equation}
and find it converges into $H$ \eqref{eq:Ham} (up to some constant) in the limit of $a_i\to 0$.  We call the eigenstates of the lattice Hamiltonian the magnetic hyperbolic Bloch states, whose detailed theoretical formulation and analysis are given in \cite{2021arXiv210710586A} 

\section{\label{sec:band}Band Structure and Quantum Gravity}
\begin{figure*}
\begin{minipage}{0.49\hsize}
        \centering
    \includegraphics[width=\hsize, bb=0 0 726 576]{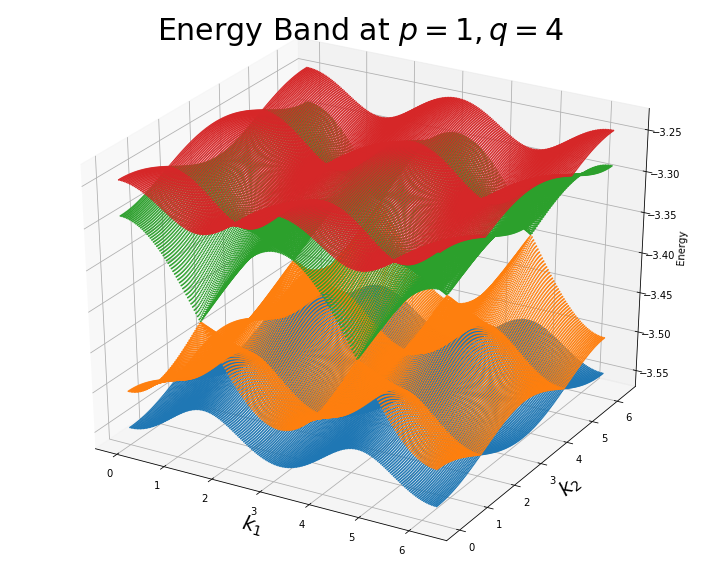}
\end{minipage}
\begin{minipage}{0.49\hsize}
        \centering
    \includegraphics[width=\hsize, bb=0 0 726 576]{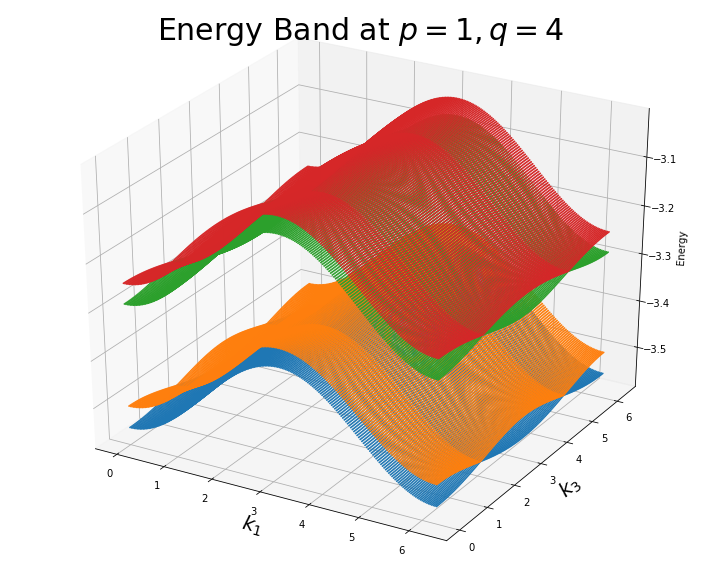}
\end{minipage}
\caption{\label{fig:band}Band structure of the electrons on the $\{8,8\}$ hyperbolic tiling. Unspecified wave numbers are set to 0. The figures correspond to some lower energy bands at $B=1/8$.}
\end{figure*}

Here we show some band structures in Fig.~\ref{fig:band}. We use a rational magnetic field $B=p/2q$, by which the energy bands split into $8q$. In general there are 8 stacks of $q$ bands, as shown in Fig.~\ref{fig:the spectrum of H lattice}. Clearly there are some band crossing points and Dirac cones, which implies the existence of a massless particle obeying the Dirac equation on a coordinate neighborhood. It is known that the peculiar electronic states (linear dispersion and gapless states) exhibited by the Dirac electron system appears on various materials: graphene, an organic conductor with a two-dimensional layered structure~\cite{PhysRevLett.102.176403}, and topological insulators. Dirac electrons have been widely used in field effect transistor devices due to its high carrier mobility in solids. Many substances showing the Dirac electron system have been explored for both experimentally and theoretically. Our work presents for the first time that a cQED system with a hyperbolic graph structure also exhibits the Diac cones. Therefore we expect that the range of materials in which the massless Dirac states appear will expand, and the development of new Dirac systems will progress. For example it will be of interest to study connections with a symmetry protected topological (SPT) phase~\cite{2017arXiv171202952M}. 

Besides the interests from material physics, let us explain some impact of our results from a view point of high energy physics as follows. As our geometry has a negative curvature, the particle should have some information of a quantum gravity. In fact a hyperbolic tiling has desirable symmetries for constructing a toy model of a gravity theory, known as the AdS/CFT correspondence~\cite{1999IJTP...38.1113M}. For example the three-dimensional AdS space is a product of a hyperbolic disc and one-dimensional time. As pointed out in \cite{2015JHEP...06..149P}, they are discretely scale-invariant, and there exist graph isomorphisms that map any point in the graph to the center while preserving the local structure of the tiling. The AdS space is an exact solution of the Einstein equation, and we may understand an eigenstate of our Hamiltonian as a state relevant to quantum gravity. In fact we expect it to travel the hyperbolic plane as fast as light due to its massless property. Extracting some gravitational information will lead to understanding a quantum gravity theory. We directly construct such a solution for the first time. A string theoretical top down approach for linking AdS/CFT correspondence with topological materials is given in~\cite{2010PhRvD..82h6014R}, whereas bottom up approaches have not been developed yet. In fact, many known low dimensional quantum gravity analysis have been limited to some strongly chaotic models dual to the black hole physics in AdS~\cite{PhysRevLett.123.031602,2011arXiv1108.1197S}. Therefore our proposing solutions and methods will open up a new direction for investigating quantum gravity solutions in a laboratory.  

\section{\label{sec:butt}Hofstadter Problem on Hyperbolic Tiling}
\begin{figure*}
\centering
\includegraphics[width=0.8\hsize, bb=0 0 1133 1700]{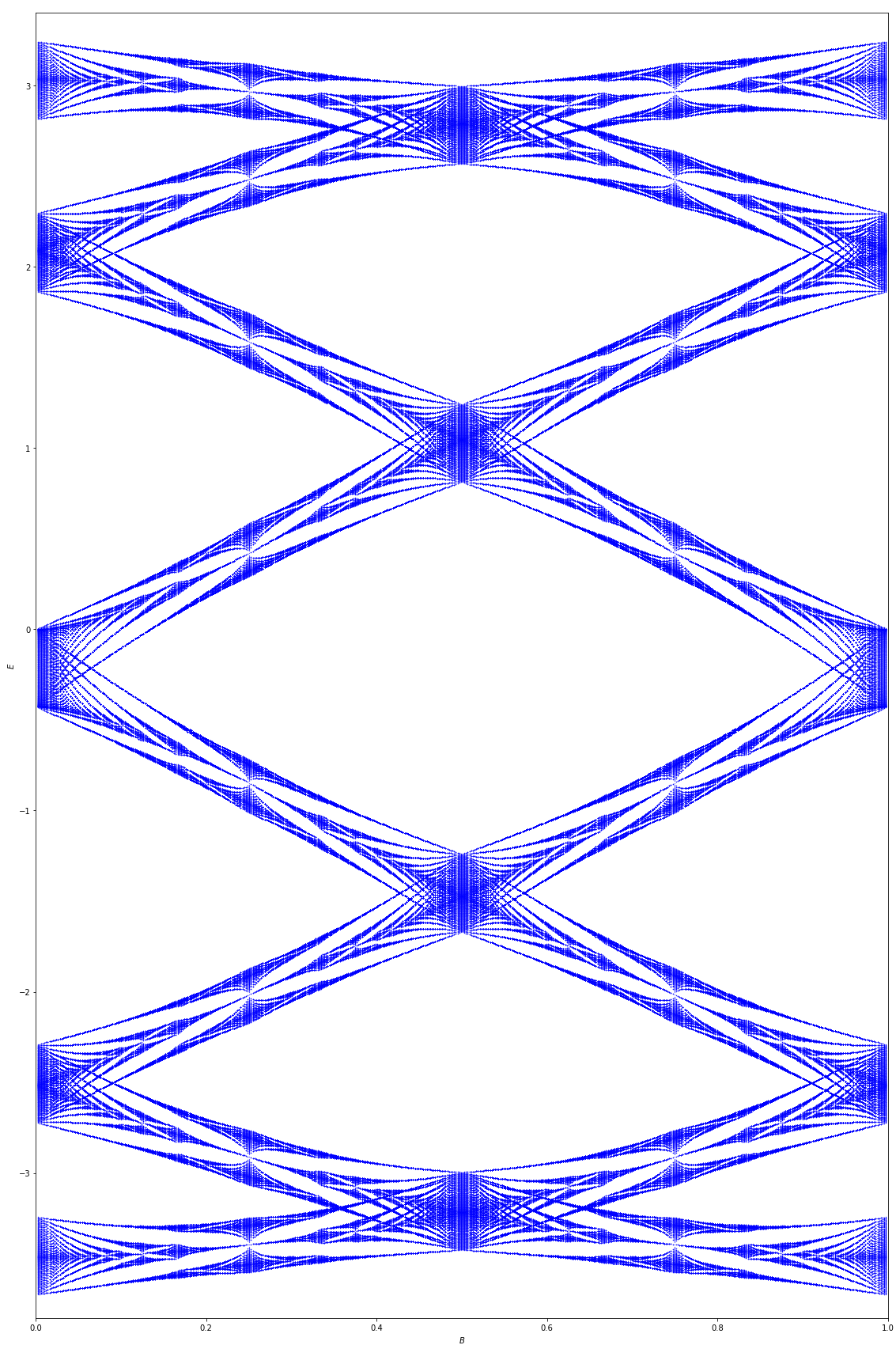}
\caption{Spectrum of an electron on a hyperbolic surface $(g=2)$ under a uniform magnetic field, plotted as a function of a rational flux. The plot is generated with the lattice Hamiltonian \eqref{eq:Ham_latt}.}
\label{fig:the spectrum of H lattice}
\end{figure*}

Here we study the energy spectra of electrons under a magnetic field from another viewpoint. The Hofstadter problem~\cite{1976PhRvB..14.2239H} is a well-known and established problem on the energy spectrum of Bloch electrons moving in a two-dimensional lattice under a uniform magnetic field. As a function of magnetic flux, the energy spectra exhibit the fractal structure, called the Hofstadter butterfly. Due to such an extremely non-trivial fractal structure, many authors have considered this problem from various theoretical viewpoints 
\cite{PhysRevB.46.12606,PhysRevLett.86.1062,Wang1231,PhysRevLett.125.236804,doi:10.1063/1.4998635,Hatsuda:2016mdw}.
Many experimental evidences of the fractal structure of the Hofstadter spectrum were reported in a lot of systems, such as GaAs-AlGaAs heterostructures with superlattices \cite{Schl_sser_1996,PhysRevLett.86.147,geisler2004detection}, ultracold atoms in optical lattices \cite{Jaksch_2003,PhysRevLett.111.185301,PhysRevLett.111.185302}, moir\'{e} superlattices \cite{moire,Hunt1427,Ponomarenko2013}, the superconducting qubits \cite{Roushan1175} 
and 1D acoustic array \cite{PhysRevLett.80.3232,Richoux_2002,ni}. In Fig.~\ref{fig:the spectrum of H lattice}, we present our solution of the Hofstadter problem. The figure exhibits the unusual self-similar structure. This may be expected since our theory is a natural and general extension of the conventional works on a Euclidean lattice. Much more crucial and surprising fact for us is that there are more than 1 stacks of butterflies that appear in a pattern while keeping some band gaps. We fix $q=499$ and run $p$ from 1 to $2q-1$. Then each stack of bands constitutes of $q$ eigenenergies of the Hamiltonian, thereby there are $8q$ bands in total. "8" comes from $4\times g~(g=2)$ and we expect this formula should be true in general for $g\ge2$. It is important to recall that the $g=1$ case, namely the standard torus case, is exceptional. This is based on the fact that any closed and oriented two-dimensional manifold is classified by its curvature: positive ($g=0$), zero ($g=1$) and negative ($g\ge2$). So our finding on the difference in the fractal structures clearly shows that the band theory on the hyperbolic geometry ($g\ge2$) is essentially different from that on the Euclidean geometry, even at some quantum scale.

\section{\label{sec:hall}Remarks on the Hall Conductance and the Brillouin Zone}
 Here we give a general description of the Hall conductance based on \cite{Ikeda:2018tlz,Ikeda:2017uce,2005hep.th...12172F,2007CNTP....1....1K}. A study of the Hall conductance on a hyperbolic surface is given by~\cite{1998math......4128C}. Our theory is a $U(1)$-gauge theory on a compact Riemann surface $\mathbb{H}/\Gamma\simeq \Sigma_g$ of genus $g\ge 2$, where $\Gamma\subset PSL(2,\mathbb{R})$ is the the discrete Fuchsian group associated with the tiling of Poincar\'{e} disk. The geometric Langlands correspondence is an equivalence between the derived category of quasicoherent sheaves on the local systems and the derived category 
of $D$-modules on the Picard variety $\text{Pic}=\bigcup_d \{\mathcal{L}: d= c_1(\mathcal{L})\}$ of $\Sigma_g$ classifying line bundles on $\Sigma_g$. Here $c_1(\mathcal{L})$ is the Chern number of a line bundle $\mathcal{L}$ on $\Sigma_g$.  

As pointed out by \cite{2020arXiv200805489M}, the generalized Brillouin zone $\text{Jac}(\Sigma_g)$ is the Jacobian of $\Sigma_g$ and parametrizes distinct $U(1)$-representations of $\pi_1(\Sigma_g)$. The degree zero component $\text{Pic}_0$ of $\text{Pic}$ corresponds to $\text{Jac}(\Sigma_g)$. In general the Hall conductance is given by the sum of the Chern number of line bundles $\mathcal{L}_i\in \text{Pic}$ below the Landau level     
\begin{equation}
    \sigma_{xy}=\sum_{i}c_1(\mathcal{L}_i). 
\end{equation}

We consider the Abel-Jacobi map $j:\Sigma_g\to\text{Jac}(\Sigma_g)$ sending $p\in \Sigma_g$ to the line bundle for some fixed reference point $p_0\in \Sigma_g$:
\begin{equation}
j(x)=\left(\int_{p_0}^p\omega_1,\cdots,\int_{p_0}^p\omega_g\right),
\end{equation}
where $\omega_1,\cdots,\omega_g$ are bases of the space of holomorphic differentials on $\Sigma_g$. By the Abel-Jacobi there is an isomorphism as complex manifolds $\mathbb{C}^g/\Lambda\simeq\text{Jac}(\Sigma_g)$, where $\Lambda$ is the lattice spanned by the integrals of $\omega_i$'s over the one-cycles in $\Sigma_g$. In particular, $\text{Jac}(\Sigma_g)$ corresponds to a torus $T^2$ when $g=1$ (the case of the conventional theory on a Euclidean lattice). In this sense we can call $\text{Jac}(\Sigma_g)$ the generalized Brillouin zone.

Now we consider a holomorphic line bundle $\mathcal{L}$ with connection on $\Sigma_g$. A local system is a pair $L=(\mathcal{L},\nabla)$ of holomorphic line bundle $\mathcal{L}$ and a holomorphic connection $\nabla$ on $\mathcal{L}$. By the Fourier-Mukai transformation~\cite{mukai_1981}, $L=(\mathcal{L},\nabla)$ is mapped to a $\mathcal{D}$-module $F=(\mathcal{F},\tilde{\nabla})$, where $\mathcal{F}$ is a holomorphic line bundle on Jac and $\tilde{\nabla}$ is a flat holomorphic connection on $\mathcal{F}$, which is a Hecke eigensheaf with respect to $\mathcal{L}$. A Berry connection would correspond to $\tilde{\nabla}$ and the Hecke functor should be regarded as the Wilson loop given by the Berry connection. Besides we can consider a generic Abel-Jacobi map and discuss the sheaves on the other components $\text{Pic}_d$ which gives a non-zero quantized Hall conductance and respects the Hecke eigensheaf feature.

\section{Conclusion and Discussion}
In this article we considered the hyperbolic band theory in the presence of a magnetic field. We analyzed the energy spectrum of the magnetic hyperbolic Bloch states (Sec.~\ref{sec:band}) and investigated the Hofstadter problem (Sec.~\ref{sec:butt}). Our proposed method is a natural and general extension of the conventional band theory on a Euclidean lattice under a magnetic field to that on a hyperbolic lattice. 

For further research directions, it will be meaningful to consider the theory with impurities or defects, by which we can address the problems under more realistic conditions. In our recent study~\cite{matsuki2021fractal,matsuki2019comments}, we showed that the fractal scaling nature of the localized length of the Bloch states localizing around a defect. Studying the localized length can be a powerful method to investigate the fractal nature of the Hofstadter butterfly as well as the energy spectrum. As our previous methods are generally true in Euclidean lattices, we expect the same logic is applicable to the hyperbolic cases. Furthermore, by identifying the boundaries by tessellation $\Gamma$, the current has some isolated zeros, due to the Poincar\'{e}-Hopf theorem \cite{milnor1997topology}. This fact is different from the conventional band theory on a square lattice or a tours, where current without an isolated zero can exits. So investigating some physical meaning of a vortex associated with such an isolated zero will be meaningful. Besides it will be yet another interesting problem to complete the classification table of topological materials on hyperbolic plane, as did for the case of the Euclidean band theory \cite{2008PhRvB..78s5125S,2017NatCo...8...50P,PhysRevX.7.041069}.

\section*{Acknowledgements}
 K.I. thanks Joseph Maciejko and Steven Rayan for useful communication. This work was supported by This work was supported by PIMS Postdoctoral Fellowship Award and by Japan Society for the Promotion of Science (JSPS) Grant-in-Aid for JSPS Research Fellow, No. JP19J20559.

\bibliographystyle{apsrev-nourl}
\bibliography{ref}

\end{document}